\documentstyle[12pt,epsf]{article}
\newcommand{\ds }{\displaystyle}
\newcommand{\ra}{\rightarrow}
\newcommand{\be}{\begin{equation}}
\newcommand{\ee}{\end{equation}}
\newcommand{\bea}{\begin{eqnarray}}
\newcommand{\eea}{\end{eqnarray}}
\newcommand{\ci}{\cite}
\newcommand{\bi}{\bibitem}
\newcommand{\nono}{\nonumber \\}

\newcommand{\dd}{\partial}

\newcommand{\s}{\sigma}

\def\dal{\,\lower0.3ex\vbox{\hrule\hbox{\vrule\kern2pt\vbox{\kern4pt\kern4pt}
\kern2pt\vrule}\hrule}\,}

\def\s{\sigma}

\begin{document}

\title{\sl Wavepackets falling under a mirror}
\vspace{1 true cm}
\author{G. K\"albermann$^*$
\\Soil and Water dept., Faculty of
Agriculture, Rehovot 76100, Israel}
\maketitle

\begin{abstract}

We depict and analyze a new effect for wavepackets falling freely under 
a barrier or well.
The effect appears for wavepackets whose initial spread is smaller
than the combination $\ds \sqrt{\frac{l_g^3}{|z_0|}}$, 
between the gravitational length scale 
$\ds l_g~=~\frac{1}{(2~m^2~g)^{1/3}}$ and the initial location of the packet
$z_0$.
It consists of a diffractive structure that 
is generated by the falling and spreading wavepacket and the waves 
reflected from the obstacle.

The effect is enhanced when the Gross-Pitaevskii interaction for
positive scattering length is included.

The theoretical analysis reproduces the essential features of the effect.
Experiments emanating from the findings are proposed.

\end{abstract}

{\bf PACS} 03.65.Nk, 42.25.Fx, 0.75.F\\

$^*${\sl e-mail address: hope@vms.huji.ac.il}

\newpage

\section{\sl Introduction}

In the past years we have described a new phenomenon called :
{\sl Wavepacket diffraction in space and time}.\ci{k1}-\ci{k5}.
The phenomenon occurs in wavepacket potential scattering for the
 nonrelativistic Schr\"odinger equation and for the relativistic Dirac
equation.
The effect consists in the production of
a multiple peak structure that travels in space and persists. This pattern was
interpreted in terms of the  
interference between the incoming spreading wavepacket and the scattered 
 wave.
The patterns are produced by a time independent potential in the
backward direction, in one dimension, and, at large angles, in three
dimensions.
The multiple-peak wave train exists for all packets, but, it does not
decay only for
packets that are initially thinner than $\ds \sqrt{\frac{w}{q}}$, where
 {\sl w} is a typical potential range or well width and {\sl q}
 is the incoming average packet momentum.
For packets that do not obey this condition the peak structure
eventually merges into a single peak.
The effect appears also in forward and backward scattering of wave packets 
from slits.\ci{k5}

The experimental breakthrough of Bose-Einstein condensation in
 clusters of alkali atoms\ci{ander,davis}
\footnote{A comprehensive bibliography
on Bose-Einstein condensation may be found at the JILA site 
http://bec01.phy.GaSoU.edu/bec.html/bibliography.html} 
lead to the reinvestigation of the influence 
 of the earth's gravitational field on the development of a quantum system.

Gravity is currently being advocated as a mean to allow the
extraction of  atoms from the condensate for the
realization of an atom laser continuous output coupler
.\ci{mewes,bloch,gerbier}
Despite the weakness of the gravitational force on earth, it has a major
influence on atoms that are cooled to microKelvin temperatures.
It is then necessary to include the effects of gravity 
in theoretical calculations with condensates.

Many other gravitational effects with quantum systems are being considered
nowadays, 
such as bound states of neutrons in a gravitational field above a mirror, 
\ci{nesvi}, or the use of the coherence properties of condensates to
serve as interferometers in the presence of gravity\ci{bongs,peters}.

The Bose-Einstein condensate in a magnetic trap is in reality a wave packet.
To the extent that decoherence effects are not dominant, it is expected to 
evolve in a gravitational field in the same manner as a Schr\"odinger
wave packet.
In the present work we investigate the effects of gravity on
falling packets.

It will be shown numerically and analytically that packets falling {\sl under}
an obstacle but, free from below, display distinctive quantum
features due to their wave nature.

The Schr\"odinger wavepackets not only fall, in accordance with the
equivalence principle, but also spread. The thinner the initial extent
of the packet the broader the spectrum of momenta it carries.
Consequently, it will generate many more components able to reflect
from the obstacle, be it a well or a barrier. These reflected waves
will interact with the spreading and falling packet.

There is crossover length scale, at which
the interfering pieces start to produce a diffractive coherent
structure that travels in time, analogous to the effect of wave packet 
diffraction in space and time previously investigated.\ci{k1}-\ci{k5}

This length scale is the gravitational scale 
$\ds l_g~=~\frac{1}{(2~m^2~g)^{1/3}}$.
For Sodium atoms it is about 0.73 microns, while
for Hydrogen it is 5.86 microns.\footnote{
We use $\ds \hbar~=~1,~c~=~1$, and units of length in microns,
of time in milliseconds, $\ds g~=~9.8 \frac{\mu}{(msec)^2}$.}
For packets {\sl initially} narrower than a proportion of this scale, 
the effect
is extremely evident, and gets blurred the wider the initial packet.
This effect may be observed with the same setup as the one used in
Bose-Einstein condensation experiments, provided pencil-like, 
thin packets are produced and allowed to fall under a roof.\ci{bongs}

We will provide analytical approximations to the exact solution of
the problem that reproduce quite satisfactorily the numerical results.
In section 2 we present numerical results.
Section 3 will deal with theoretical aspects.
Section 4 summarizes the paper and provides concluding remarks
regarding possible experiments.

\section{\sl Packets falling under a roof}

Matter wave diffraction phenomena in time \ci{mosh} induced by the sudden
opening of a slit, or in space by fixed slits or gratings,
are understood simply by resorting to plane wave monochromatic waves.

Atomic wave diffraction experiments \ci{prl}, have confirmed the predictions
of diffraction in time\ci{mosh} calculations.
These patterns fade out as time progresses.

The phenomenon of diffraction of wavepackets 
in space and time was presented in \ci{k1}-\ci{k5}.
It consists of a multiple peak traveling structure 
generated by the scattering of initially thin packets from 
a time independent potential, a well, a barrier, or a grating.
The condition for the pattern to persist was found to be

\bea\label{constraint}
\s<<\sqrt{\frac{w}{q_0}}
\eea

where $\ds \s$ is the initial spread of the packet, $\ds w$ is the width of the well 
or barrier and $\ds q_0$ is the impinging packet average momentum.
For packets broader than this scale the diffraction pattern mingles into
a single broad peak.

The original motivation for the present work, was the addition of gravity
to the potential affecting the packet propagation.
As described above, the effects of gravity become increasingly
relevant to the dynamics of packets in traps and elsewhere.

The educational literature abounds in works dealing with the 
dynamics of packets falling on a mirror. The so-called {\sl quantum bouncer}
\ci{bana} is a clean example of the use of the Airy packet in the 
treatment of the problem. The use of the Airy packet is straightforward
above the mirror with the boundary condition of a vanishing wave function
at the location of the mirror, and, becomes a nice laboratory for
the investigation of the quantum classical correspondence, revivals, the Talbot
effect, etc.
Falling packets were not studied, perhaps in light of the preconception
 that nothing
interesting will be found besides the expected spread and 
free fall of the packet.

However, there is a surprise awaiting us here.
This is not totally unexpected due to the wave nature of the packet that
consists of modes propagating in both the downwards and the upwards
direction. Analogous effects are apparent also when a packet propagates
in parallel to a mirror without ever getting close to it.\ci{dodo}
Again the spreading and interference between the incoming and reflected
waves produce a wealth of phenomena.

In this section, 
we present the numerical results for the falling of packets under a
barrier or well using Gaussian wave packets. The use of
Gaussian packets permits a
straightforward connection to theoretical predictions.

The scattering event starts at {\sl t =0} with a minimal uncertainty wavepacket

\bea\label{packet}
\Psi_0~&=&~A~e^{\i\alpha}\nono
\alpha&=&~q~(z-z_0)-\frac{(z-z_0)^2}{4~\s^2}
\eea

centered at a location $\ds z_0$ large enough for the packet to be
almost entirely outside the range of the potential.
$\s$ denotes the width parameter of the packet.
$ q = mv$ is the average momentum of the packet.
The potential affecting the packet is a square well, the gravitational
interaction, and eventually the Gross-Pitaevskii (GP) 
interaction, that subsums the effects of forces between the atoms 
in the condensate in the mean field approximation\ci{pita}.

\bea\label{pot}
V~=~m~g~z+~U~\Theta(w/2-|z-w/2|)+~g~|\Psi|^2
\eea

where $\ds m$ is the mass of the atom taken here to be Sodium, {\sl w} is
the width of the well or barrier, of depth or height {\sl U}, and {\sl g}
is the strength of the GP interaction\ci{pita}, which
using the scattering length of Sodium and a typical number of atoms
in a trap of the order of 250 atoms/$\mu^3$ yields $\ds g~\approx~25$ for
a wavefunction normalized to one.

Figure 1 depits the setup of the problem. A thin wave packet is initially
located around a point $z_0~=~-7~\mu$ under a thick barrier, the mirror.
The mirror could be any flat surface. We have represented the
 strength of the barrier by means of a square located above $z~=0$.
The simplification of the present investigation assumes a the mirror
to be a finite width plate of thickness $ t~= 10 \mu$ in the
figure extending to a much larger distance along the {\sl x,y}
plane.
For the one dimensional calculation we took a thin packet along the
{\sl z axis} of width $4\s~\approx 0.4\mu$. It is supposed to be a pencil-like
packet whose extension along the {\sl x,y} plane is larger
than the width , such that the quantum dispersion affects primordially
 the behavior of the packet along the vertical direction. In this
direction, the facket falls freely. The interference between
 the packet components reflected from the barrrier, upwards moving
 waves, and those that
 fall freely without reflection, the downward moving ones, will induce
 a diffraction pattern.
We will find that the contrast of the pattern depends crucially on the
initial location of the packet and its width.

An ideal implementation of this setup would be to use a Bose-Einstein 
condensate optically trapped in a region below a plate. Suddenly
stop the confining potential and let it fall freely under the plate.

The condition for the observation of the effect would be to 
produce a thin enough condensate. For repulsive
Gross-Pitaevskii interactions as in Sodium, Rubidium, or Hydrogen,
the width limitation is less stringent, as will be shown below.
The extra repulsion of the interaction,
 on top of the quantum dispersion, 
effectively pushes the waves inside the condensate against the wall.
 It acts as if the packet was initially thinner than it actually is.
In particular, for Hydrogen the initial width of the packet
may be as large a few microns.
In section 4 we make some further comments concerning the relevance
of the effect for atom lasers.

\begin{figure}
\epsffile{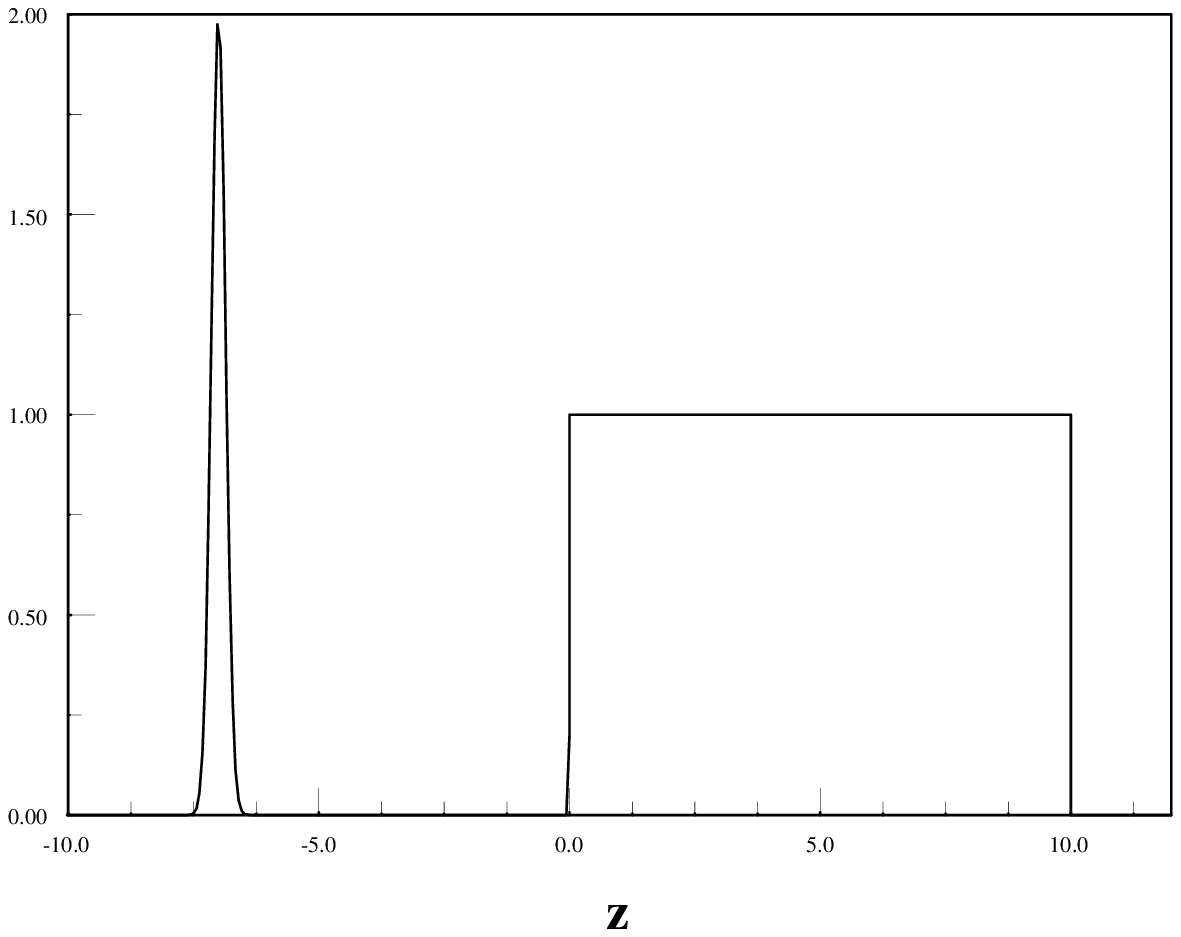}
\vsize=5 cm
\caption{\sl Setup of the problem. The packet's
initial location is $z_0=~-7\mu$. The mirror is depicted as a 
square of arbitrary height representing the strength of the
 repulsive potential. The mirror is located above $z~=0$ and its width is
10$\mu$}
\label{fig1}
\end{figure}

The algorithm for the numerical integration of the 
Schr\"odinger equation of the present work  
is described in previous works.\ci{k1,k2,k3}.
Flux conservation for initially normalized packets 
and energy conservation require that

\bea\label{flux}
1&=&\int_{-\infty}^{\infty}~|\Psi(z,t)|^2~dz\nono
E&=&\int_{-\infty}^{\infty}\bigg[\frac{1}{2~m}\frac{\dd \Psi(z,t)}
{\dd z}\frac{\dd \Psi^*(z,t)}{\dd z}~+(~m~g~z+~V(z)+\frac{g}{2}~|\Psi(z,t)|^2
)~|\Psi(z,t)|^2\bigg]~dz
\eea
with {\sl E} a constant independent of time.

The numerical runs presented below achieved an accuracy  
in the flux conservation of around 0.2\%, while
the accuracy in the energy conservation was around 2\%.
The wave function was also found to obey the Schr\"odinger equation to
an accuracy better than the 1\% level even at points near the edges
of the integration range. 

The only length scale appearing in the problem is easily derived from
the Schr\"odinger equation to be $\ds l_g$.
For Sodium this is $\ds l_g~=~0.73 \mu$

We consider the free fall of packets with initial widths $\ds \s$ smaller
and larger than a fraction of $\ds l_g$ with or without the GP interaction.
We take the barrier to have the fixed strength
of $\ds U~=10^6 sec^{-1}$ obtained from $\ds U\approx~=~\frac{4~\pi~a}{m}~N$,
with {\sl a}, the scattering length of the Sodium-solid
scattering and {\sl N} the density of a typical solid.
For the well we use a value taken from a van der Waals type of strength
\ci{grisen} at a distance of 1 nm, namely $\ds U\approx - 10^2 sec^{-1}$.
We used a very high value for the attractive well strength, beyond
the limit of applicability of the Lennard-Jones formula\ci{grisen}, 
to see whether even a large and unrealistic value for the attractive
potentials influences the results as compared to the repulsive case.

Figure 2 shows the time evolution of a thin packet falling under
a mirror. This is essentially what is expected to be
 observed in an actual imaging of a Bose-Einstein
condensate falling freely under a mirror as a
function of time. The figure shows the results for Sodium, however,
as mentioned above the same will occur with a Hydrogen condensate,
with much wider packets initially.

\begin{figure}
\epsffile{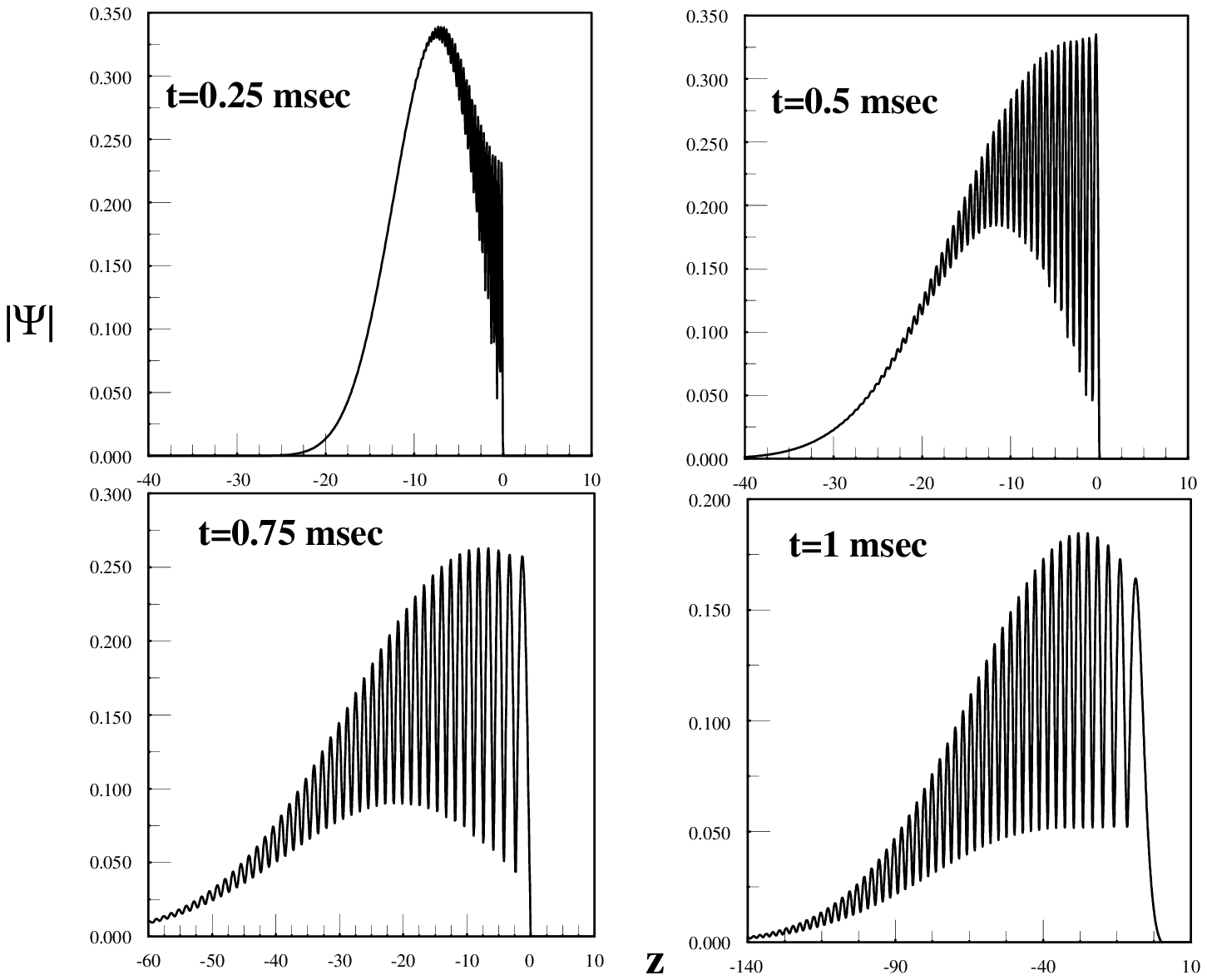}
\vsize=5 cm
\caption{\sl Time evolution of a packet profiles of initial 
location $\ds z_0~=~-7~\mu$ falling 
under the repulsive barrier (not depicted).}
\label{fig2}
\end{figure}

Figure 3 shows wide and thin packets profiles after 4 msec fall under 
a repulsive barrier. 

\begin{figure}
\epsffile{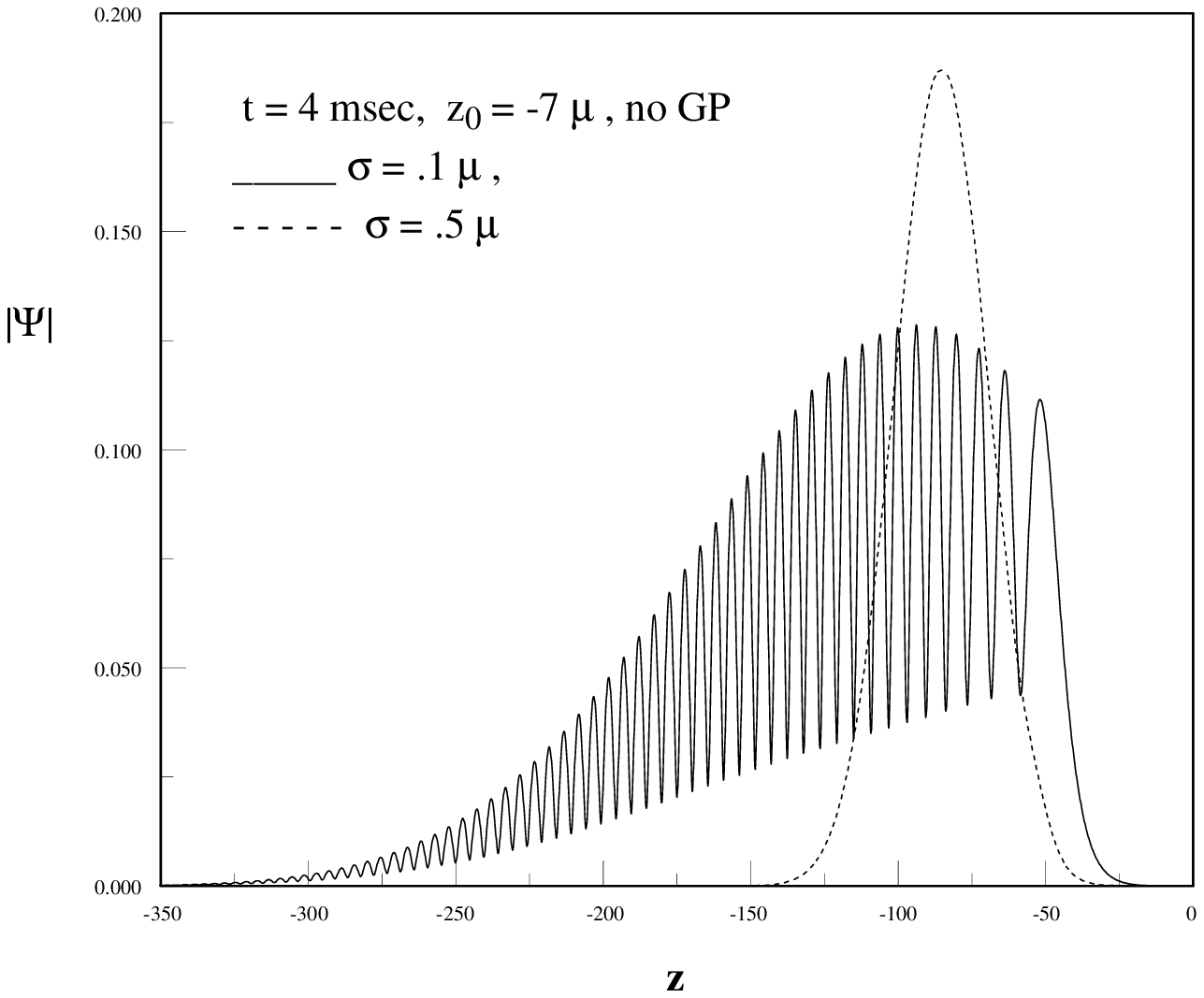}
\vsize=5 cm
\caption{\sl Packets profiles of initial 
location $\ds z_0~=~-7~\mu$ after t= 4 msec falling 
under a repulsive barrier.}
\label{fig3}
\end{figure}

The behavior of a thin packet is qualitatively different. 
A wide packet falls undistorted except for the natural spreading.
A thin packet whose width is smaller than 
$\sqrt{\frac{l_g^3}{z_0}}$ (see below for the appearance of
$z_0$ in the expression), possesses a distinctive 
diffractive structure.
The rightmost (upper) edge of the structure resembles the Airy packet
absolute value, however, the packet drops exponentially
at large $\ds |z|$ values, whereas the Airy packet diminishes
as $\ds |z|^{-1/4}$.
In the next section we will address a theoretical approach to
the problem. Approximate analytical solutions will
be provided that reproduce the basic features of both the thin and wide
packets.

In figure 4 we present an analogous picture for the case including
the GP interaction.
As expected\ci{pita}, the distinctive feature of the inclusion
of this interaction is an extra repulsive force, due to the positive
scattering length, that enhances the broadening
of the packet, and consequently the diffraction effect.
As seen from the figure, for the same values of initial packet widths,
the nonlinear repulsive interaction produces a much cleaner
interference pattern than the case lacking it in figure 3. 

\begin{figure}
\epsffile{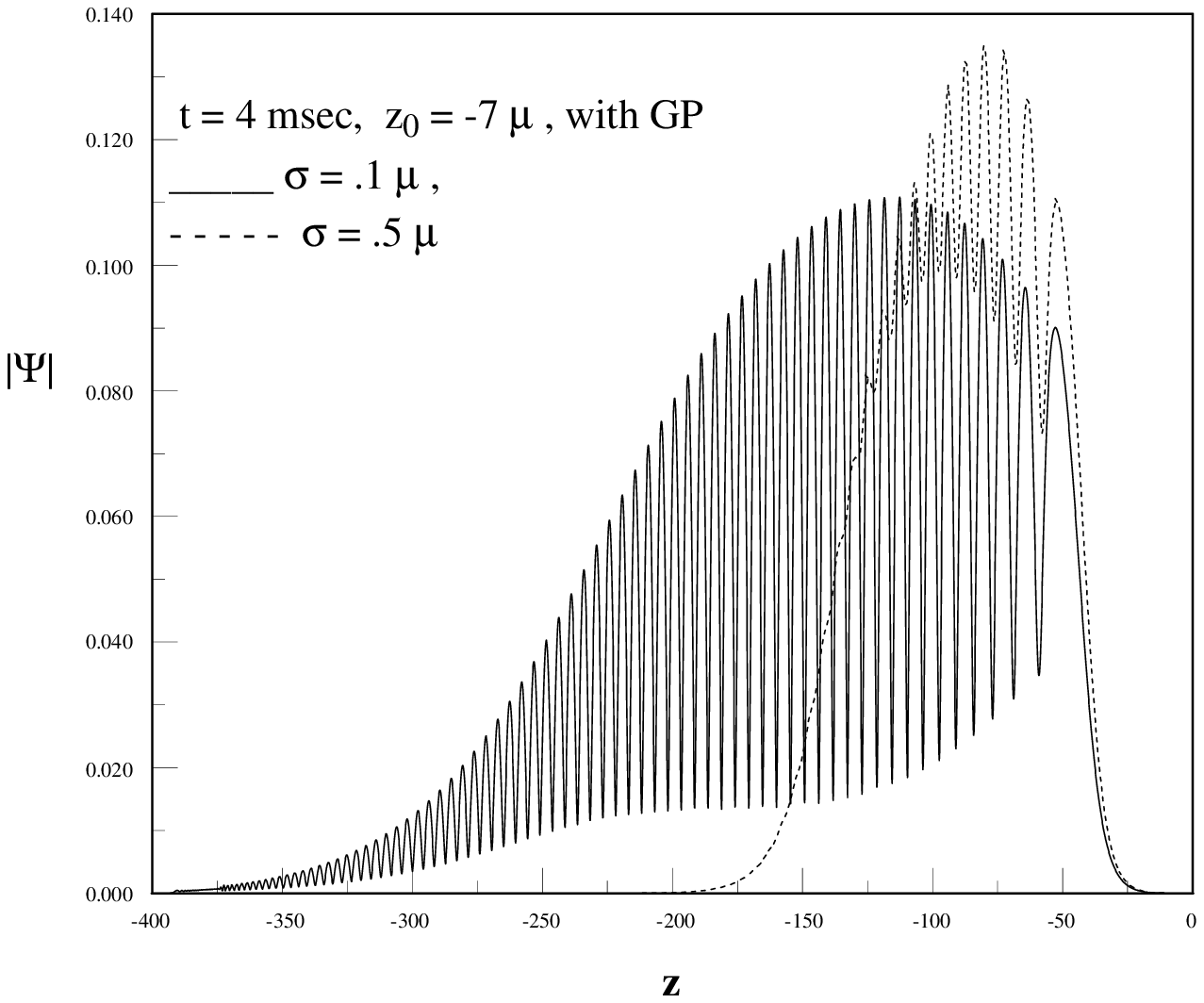}
\vsize=5 cm
\caption{\sl Packets profiles of initial 
location $\ds z_0~=~-7~\mu$ after t= 4 msec falling 
under a repulsive barrier and subjected to the Gross-Pitaevskii
interaction}
\label{fig4}
\end{figure}

Figure 5 depicts the influence of the GP interaction on the
packets profiles for initially thin packets.
 
The GP force produces a diffractive structure that has a much 
starker contrast. Large momenta are excited by the repulsive interaction,
 producing a more effective superposition between incoming and
reflected waves.

\begin{figure}
\epsffile{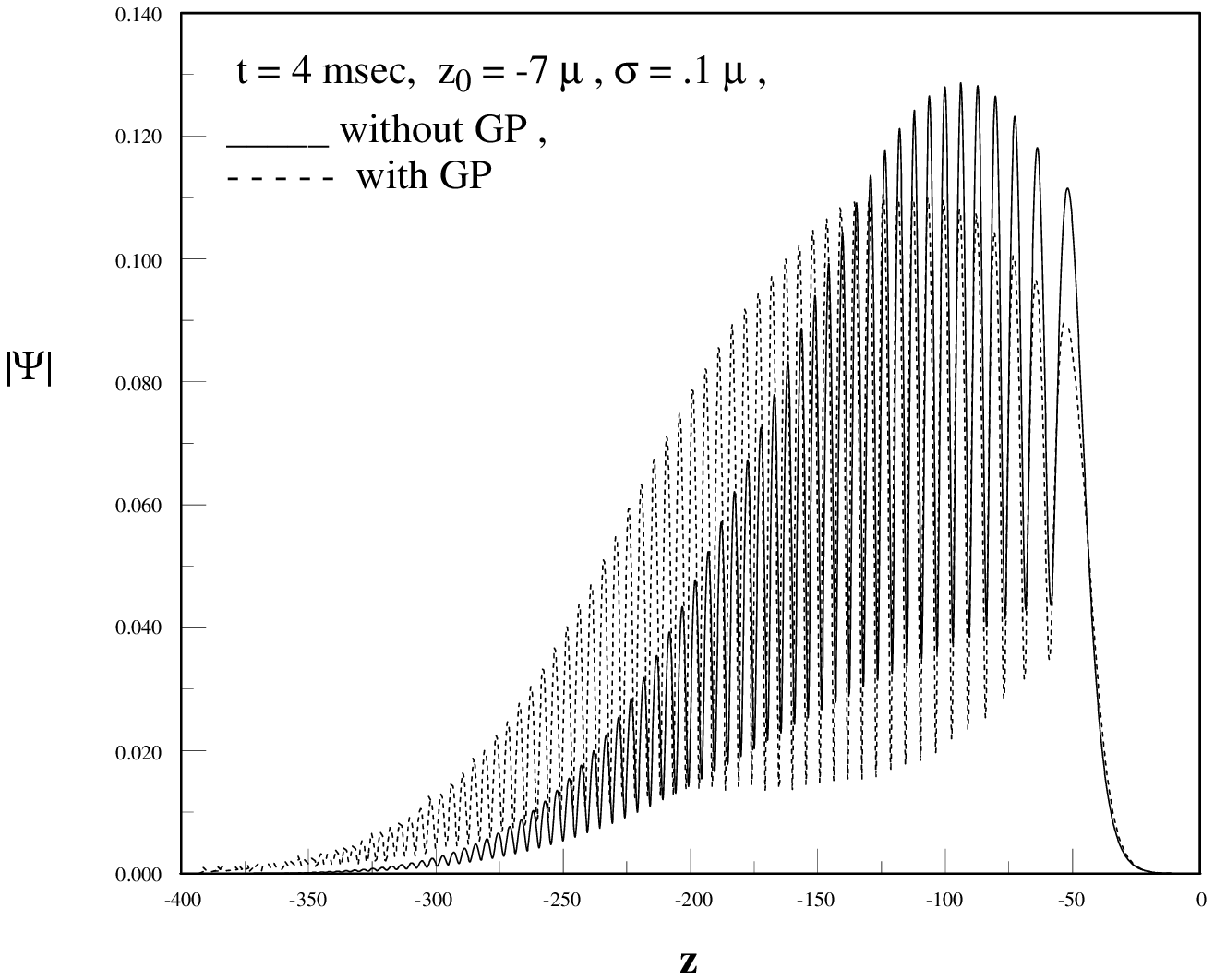}
\vsize=5 cm
\caption{\sl Thin packets profiles after t= 4 msec of fall 
under a repulsive barrier with and without the Gross-Pitaevskii interaction.}
\label{fig5}
\end{figure}
\begin{figure}
\epsffile{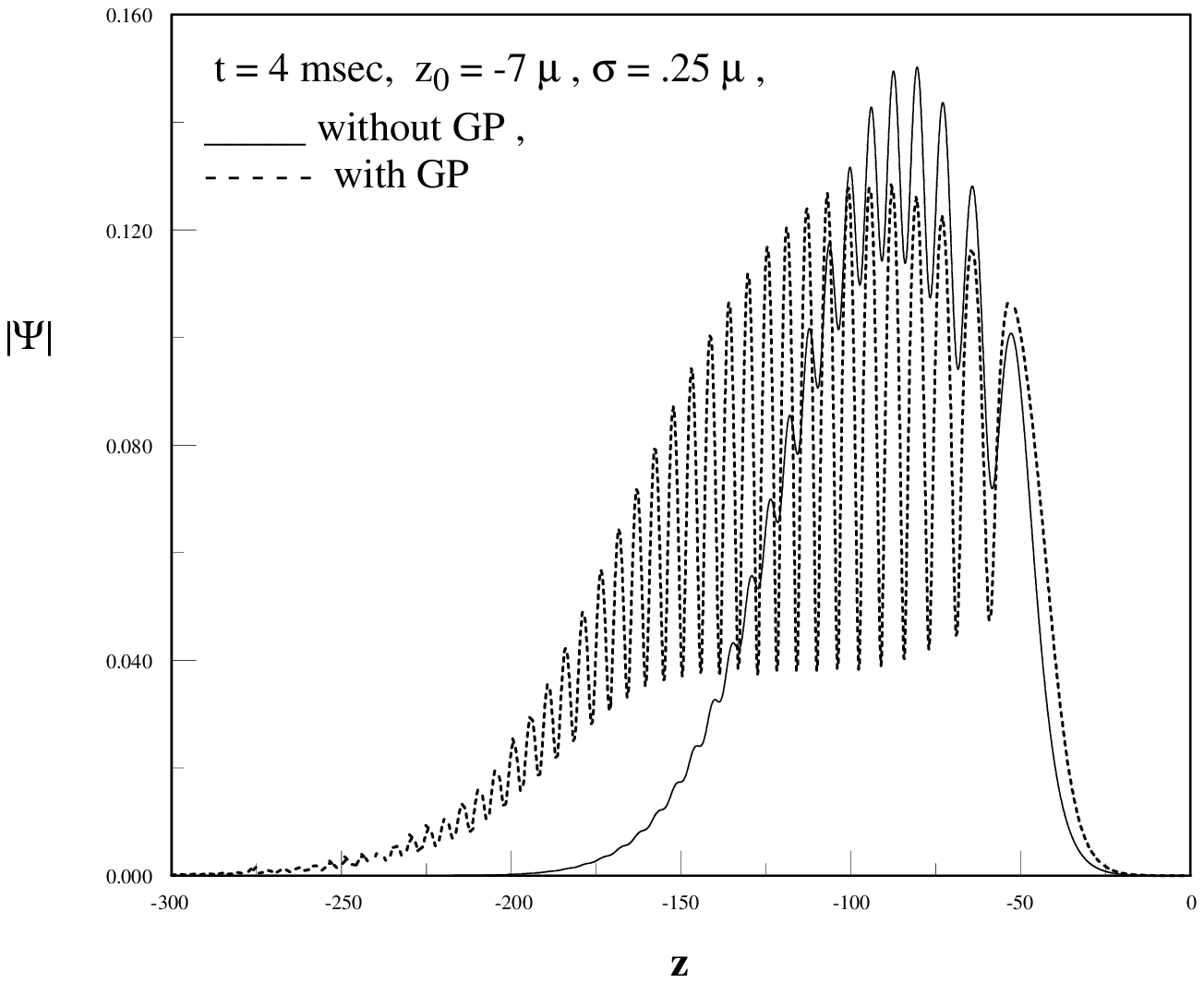}
\vsize=5 cm
\caption{\sl Borderline packets profile after t = 4 msec of fall 
under a repulsive barrier without and 
including the Gross-Pitaevskii interaction.}
\label{fig6}
\end{figure}

We argued that the appearance of a diffractive structure 
is determined by the ratio 
$\ds \epsilon~=~\frac{\s}{\sqrt{\frac{l_g^3}{z_0}}}$. 
Figure 6 shows the results
when this ratio is around 1. While the pattern has almost disappeared
from the falling packet devoid of GP interaction, it has lost contrast,
but not disappeared completely from the packet subjected to the GP force.
$\epsilon$ is then the relevant ratio for the appearance of the diffractive 
structure.

Replacing the barrier by a well has no effect whatsoever for packets without 
initial momentum. The higher the initial momentum of the packet (if positive),
the easyer the transmission through the well. The differences arise
then for packets thrown against the obstacle only at large momenta as
compared to $\ds l_g^{-1}$.
This point will be dealt with in the future.

\section{\sl Theoretical approach to falling packets}

The Airy function \ci{landau} is the solution of the
Schr\"odinger equation for the propagation in a uniform field. 
Despite being named
Airy packet, it is not normalizable and belongs to 
the set of wave functions in the continuum.
It is the analog of a plane wave in free space.

The educational literature abounds in references to uses of this packet, both
in the context of the {\sl quantum bouncer}\ci{bana} and in the treatment
of generalized Galilean transformations.\ci{green}

For the solution of a packet falling on a mirror, the {\sl quantum bouncer}, 
the Airy functions shifted to positions that have a zero at $\ds z~=~0$, serve
as a basis to find the time development of an arbitrary initial packet.
The non-normalizability of the Airy packet is of no hindrance here, because
only the upper decaying part of the packet is used. Note however that
 an aspect that is ignored in the literature is the absence of orthogonality
between the different shifted packets when integrated only
over the positive {\sl z} axis. This is perhaps not a severe
problem, but was not taken into account in the works using the 
Airy packet for the {\sl quantum bouncer} problem.\ci{bana}

The lower piece of the Airy packet is oscillatory and does not 
decay fast enough to 
serve as a basis for packets initially located under the mirror.
Only when a continuum of energies (both positive and
negative) is used, it is possible to
expand an initial wavepacket located at {\sl z} $\ds <$ 0 in terms of 
Airy functions.

Before analyzing the problem by means of Airy functions, we will develop
a very simple approximation that allows the identification of
the relevant length scale for the effect to occur.

We will resort to a set of solutions that connects
directly to plane waves. This solutions are not stationary in the 
rigorous sense of the word, because they have a time dependent phase
that is not linear in time.
The solutions are derived by transforming a plane wave to
an accelerating frame \ci{green,wadati,vande}, namely

\bea\label{sols}
{\chi}_k(z,t)&=&e^{i~\phi}\nono
\phi&=&~-m~g~z~t+~k~(~z+~\frac{g~t^2}{2})~
-~\frac{k^2}{2m}~t-\frac{m~g^2~t^3}{6}
\eea

Direct substitution in the Schr\"odinger equation proves that this
family of solutions solves the equation for a potential $\ds V~=~m~g~z$.

Given that the initial wavepacket of eq.(\ref{packet}) may be expanded
readily in plane waves that coincide with eq.(\ref{sols}) at {\sl t=0},
the Schr\"odinger equation then insures that the subsequent propagation
of the packet will be obtained by replacing the plane waves by the
solutions in the gravitational potential of eq.(\ref{sols}).

We find the expression at all times for a freely falling normalized
gaussian packet

\bea\label{packt}
\Psi(z-z_0,t)~&=&\frac{e^{\xi}~\chi_q(z-z_0,t)~\sqrt{\s}}{\sqrt{\sqrt{2~\pi}
~\s^2(t)}}\nono
\xi &=&\frac{(z-z_0+g~t^2/2-~q/m~t)^2}{4~\s^2(t)}\nono
\s^2(t)&=&\frac{i~t}{2~m}+\s^2
\eea

Figure 7 depicts the comparison between the expression of eq.(\ref{packt})
and a numerical calculation. The agreement is satisfactory without
any rescaling. It lends confidence in both the numerical scheme and the
theoretical formulae.

\begin{figure}
\epsffile{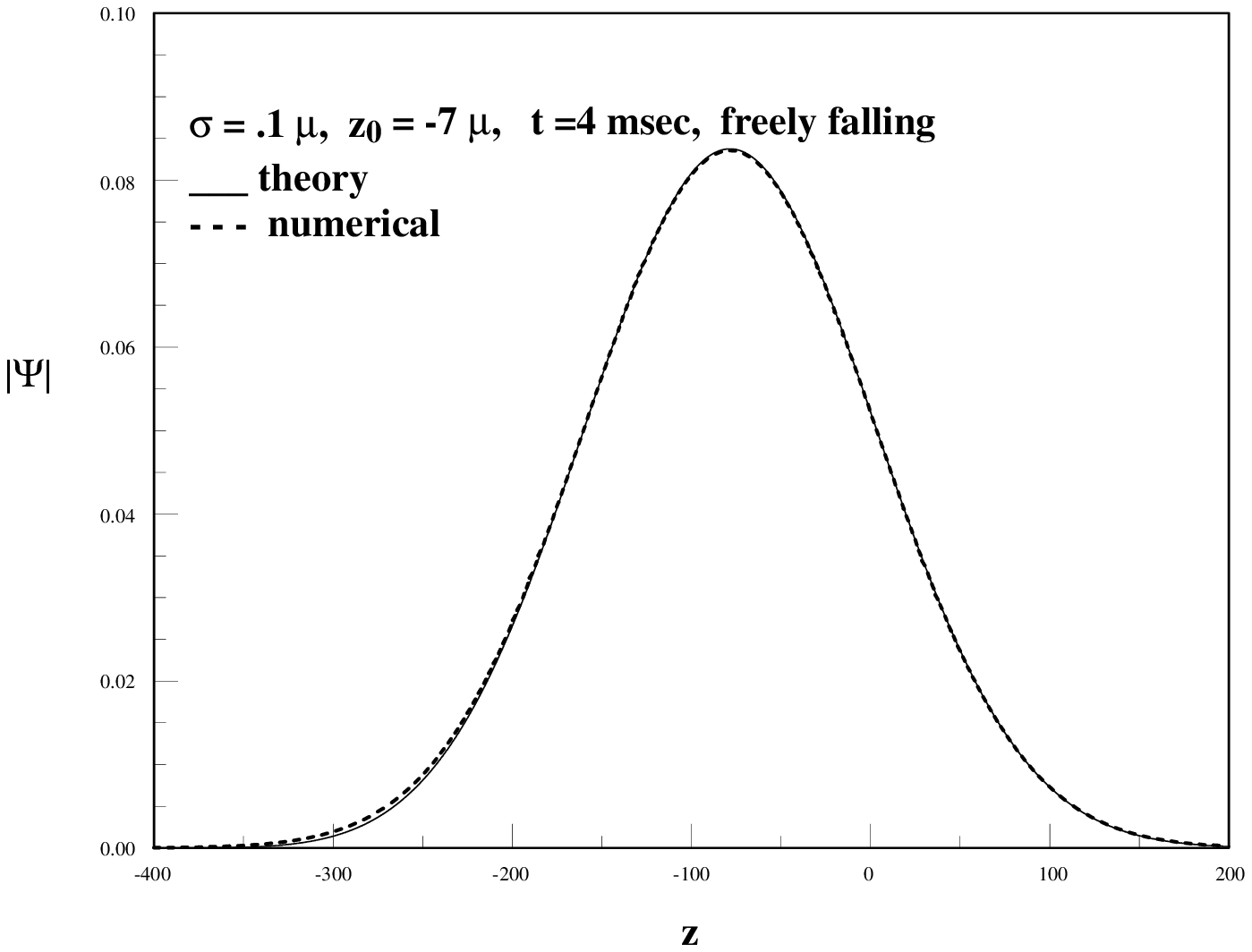}
\vsize=5 cm
\caption{\sl Profile of a packet of initial width $\ds \s=0.35~\mu$ and initial
location $\ds z_0~=~-7~\mu$ in free fall after t = 4 msec. 
Numerical calculation, dotted line, and theoretical formula of eq.(\ref{packt})
, full line}
\label{fig7}
\end{figure}

At {\sl t=0} we can readily build a packet that solves the
problem with the boundary condition of $\ds \Phi(0)=0$, 
corresponding to an impenetrable mirror at {\sl z~=~0}.
Just an image packet located at $\ds -z_0$ in the inaccessible region
above the mirror will serve.
The solution at {\sl t=0} then becomes 
$\Phi(z,t=0)~=~\Psi(z-z_0,0)-\Psi(z+z_0,0)$, with $\ds \Psi$ in eq.(\ref{packt}).
$\Phi$ obeys the equation of motion 
and the boundary condition. It also coincides with the initial packet of 
eq.(\ref{packet}) in the allowed region of {\sl z} $\ds <$ 0.
Propagating $\ds \Phi$ forward in time we obtain

\bea\label{pckt1}
\Phi(z,t)\approx\Psi(z-z_0,t)-\Psi(z+z_0,t)
\eea

We write the approximate sign, because the cancellation at {\sl z~=~0} is
only effective at short times. As soon as {\sl t} increases, the wave
function of eq.(\ref{pckt1}) does not vanish any more.
Presumably, an infinite set of image packets is needed.

We could not find a closed analytical solution at all times.
The solution of eq.(\ref{pckt1}) reproduces reasonably well the
falling packets and distinguishes clearly between a packet narrower than $\ds l_g$
 and one wider than $\ds l_g$ for $\ds z_0$ of the order of a few packet widths.
In order to see this  
we write the absolute value of eq.(\ref{pckt1}) for a packet
with initial momentum $\ds q=0$ for $t>>~2~m~\s^2$

\bea\label{abs}
|\Psi(z,t)|&=&~A~e^{\theta_1}~\sqrt{sin^2(\theta_2)+
sinh^2(\theta_3)}\nono
\theta_1&=&-\frac{m^2~\s^2~\bigg((z+g~t^2/2)^2+z_0^2\bigg)}{4~t^2}\nono
\theta_2&=&\frac{m~z_0~(z+g~t^2/2)}{t}\nono
\theta_3&=&\frac{m^2~\s^2~z_0~(z+g~t^2/2)}{2~t^2}
\eea

$\theta_3$ is responsible for the blurring and loss of contrast of the
diffraction pattern determined by the {\sl sin} function.
The larger $\ds \theta_3$ the less visible are the oscillations.
The criterion for the visibility of the pattern may then be written
as $\ds max(\theta_3)<<1$. The maximum value of {\sl - z} is given by the
descent of the packet and its spreading. Both are of the order of
$\ds \frac{g~t^2}{2}$. We can then write the condition for the visibility
of the interference fringes to be
\bea\label{fringe}
max(\theta_3)\approx\frac{m^2~\s^2~z_0~g}{2}<<1
\eea

However, typically $\ds z_0$ amounts to a few times 
the width of the packet, otherwise the oscillations will have
a very large wavenumber, and will blur the pattern anyway.

Hence we can write eq.(\ref{fringe}) as

\bea\label{fring}
max(\theta_3)&\approx&\frac{m^2~\s^3~g}{2}<<1\nono 
&=&\frac{\s^3}{4~l_g^3}<<1\nono
or\nono
\s~<<~l_g
\eea
If we keep $\ds z_0$, eq.(\ref{fring}) becomes

\bea\label{fring1}
\s~<<~2~\sqrt{\frac{l_g^3}{z_0}}
\eea

Eq.(\ref{fring1}) demonstrates that the relevant borderline
between a visible and blurred packet is $\ds \s=\sqrt{\frac{l_g^3}{z_0}}$.
A packet initially narrower that $\ds l_g$ located at a distance
of a few times its width under a mirror will definitely
display interference fringes.
Eq.(\ref{abs}) also tells us that this pattern travels with the
packet unscathed.

For long enough times we can improve upon the solution of
eq.(\ref{pckt1}) in order to compensate for the inaccuracy in the
cancellation at $\ds z~=~0$. The correction is achieved by multiplying
the subtracted wave by a space independent, but
time dependent admixture factor,

\bea\label{pckt2}
\Phi(z,t)&\approx&\Psi(z-z_0,t)-\lambda\Psi(z+z_0,t)\nono
\lambda&=&\frac{\Psi(-z_0,t)}{\Psi(z_0,t)}
\eea

By adding this factor we spoil the solution. Eq.(\ref{pckt2}) does not solve
exactly the Schr\"odinger equation, whereas eq.(\ref{pckt1}) does.
However, this inaccuracy decreases as a function of time, because $\ds \lambda
\ra~1$
as $\ds t\ra\infty$. For times not so long it improves a little the 
agreement with the numerical results. It apparently compensates for the 
need of an infinite set of packets that insure the boundary condition at all 
times.

Figures 8 and 9 show the comparison between the formula of eq.(\ref{pckt2})
and the numerical results both for a thin packet and wider one.

The agreement is reasonable quite good for the wide packet and
qualitatively reasonable for the thin packet. 
The formula captures 
the gross features, such a as the absence of a diffractive structure for
a wide packet and the wavenumber of the oscillations for
a thin packet. It has one evident limitation: Lack of a sharp
cutoff of the packet at small distances {\sl -z} for a thin packet.
Nevertheless, the simple picture of single image packet is essentially correct.

\begin{figure}
\epsffile{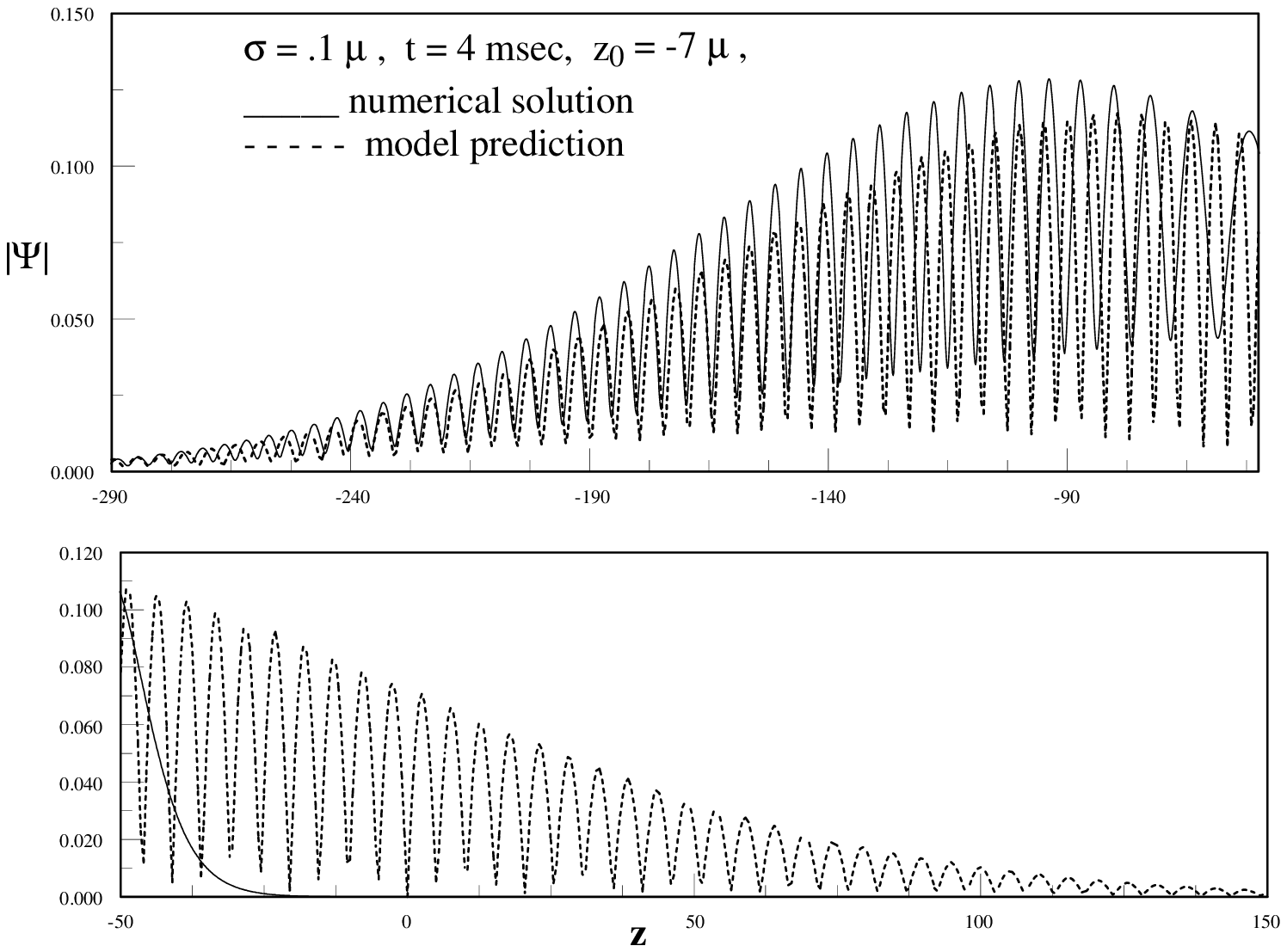}
\vsize=5 cm
\caption{\sl Thin packet profile
of width $\ds \s=0.3~\mu$ after t = 4 msec falling under a reflecting mirror. 
Numerical calculation, solid line, and theoretical formula of eq.(\ref{pckt2})
}
\label{fig8}
\end{figure}

\begin{figure}
\epsffile{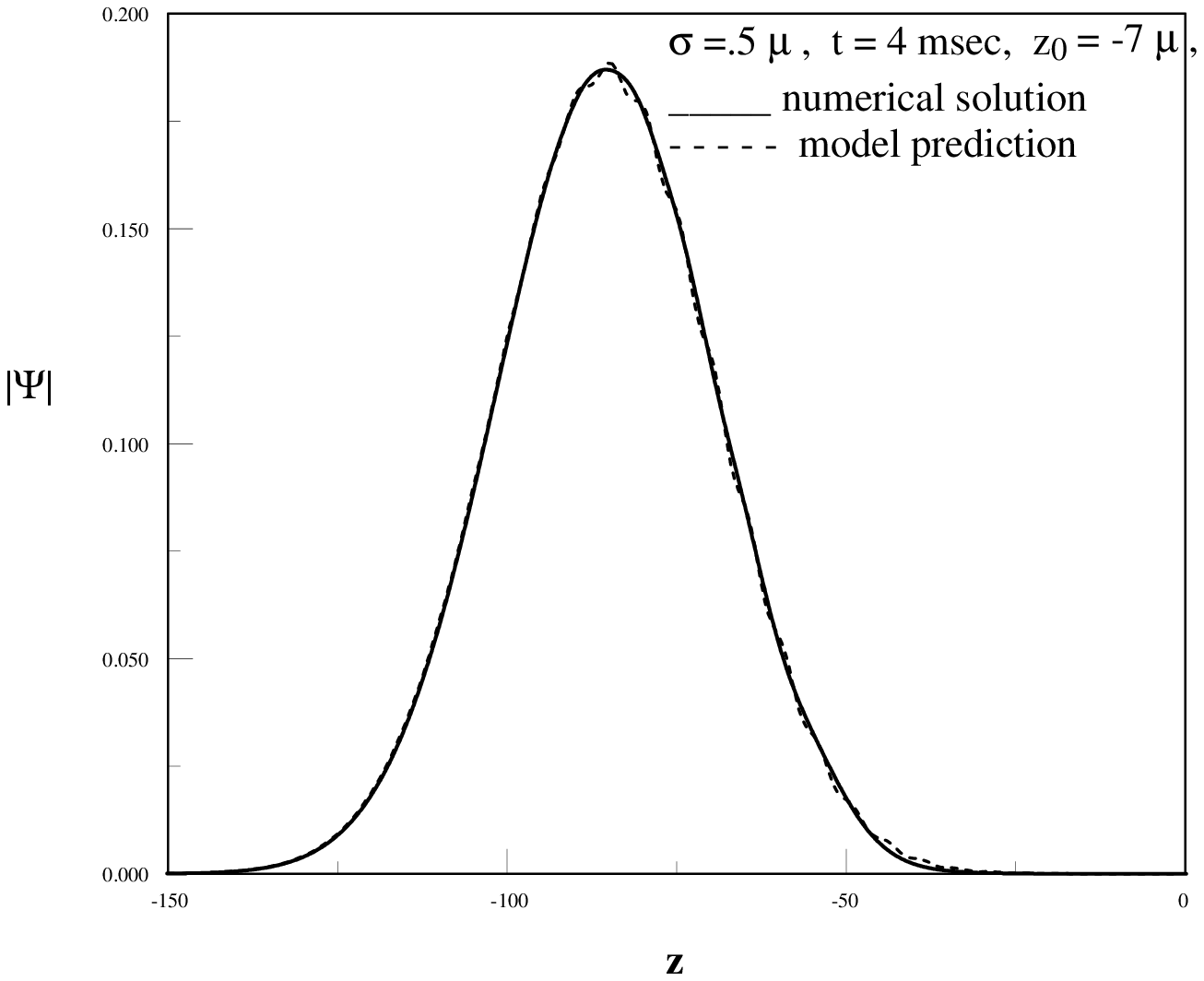}
\vsize=5 cm
\caption{\sl Wide packet profile 
of width $\ds \s=2~\mu$ after t = 4 msec falling under a reflecting mirror. 
Numerical calculation, solid line, and theoretical formula of eq.(\ref{pckt2}
)}
\label{fig9}
\end{figure}

\newpage
We now proceed to find a solution in terms of Airy functions.\footnote{The
author wishes to express his gratitude for the extremely helpful
remarks of one of the anonymous referees concerning the use of
Airy functions}

A solution of the mirror problem for $z~<~0$ can be determined
by means of a set of functions that solve the freely
falling packet, namely linear combinations of the independent
Airy functions {\sl Ai and Bi}\ci{abramo}. A set that implements the
boundary condition of a vanishing wave function at the origin reads

\bea\label{chi}
\chi_a(x)&=&(Bi(a)~Ai(x+a)-Ai(a)~Bi(x+a))/\sqrt{Ai^2(a)+Bi^2(a)}
\eea

where $\ds x~=~\frac{z}{l_g},~a~=\frac{-E}{mgl_g}$, with
{\sl E}, the energy of a stationary solution $-\infty<~a~<\infty$.

The functions obey
\bea\label{compl}
\int_{-\infty}^{0}\chi_a(x)~\chi_b(x)~dx~=~\delta(a-b)
\eea

with $\delta$, the Dirac $\delta$ function.
The set is orthonormal and complete.

The initial wave packet of eq.(\ref{packet}) is expanded in terms
of the set above as

\bea\label{expan}
\Psi(z,t=0)&=&\int_{-\infty}^{\infty}~C(a)\chi_a(x)~da\nono
C(a)&=&\int_{-\infty}^0\chi_a(x)\Psi(x,t=0)~dx
\eea

The wave function at all times then becomes
\bea\label{expant}
\Psi(x,t)&=&\int_{-\infty}^{\infty}~C(a)\chi_a(x)~e^{i~E~t}~da\nono
\eea

For thin enough wave packets $C(a)$ may be obtained analytically by
extending the integration to $+\infty$ and neglecting corrections of
order $\ds\gamma=\frac{\s}{l_g}$, with $\s$, the initial
width of the packet.
 Under these approximations C(a) becomes

\bea\label{c}
C(a)=\bigg(2^3\pi\gamma^2\bigg)^{\frac{1}{4}}~l_g~exp((a+x_0)~\gamma^2+
\frac{2~\gamma^6}{3})~\chi(a+x_0+\gamma^2)
\eea
 
where $\ds x_0=\frac{z_0}{l_g}$

Inserting this formula into eq.(\ref{expant}) we find the wave
function at all times.

Figures 10 and 11 present the comparison of the numerical
calculation for a thin packet and a wider packet to the
analytical formula of eq.(\ref{expant}) with the approximation
of eq.(\ref{c}).
The thin packet fit is quite good, while for the wider
packet there is a discrepancy in the
height of the peak. This discrepancy is due to the fact that $\gamma$ for
a wider packet is not negligible and the formula of eq.(\ref{c}) needs
to be ammended. However in both cases the basic feature
of the existence and absence of a diffractive train is evident.
It is worth mentioning that there is no rescaling of neither
analytical nor numerical data points.

\begin{figure}
\epsffile{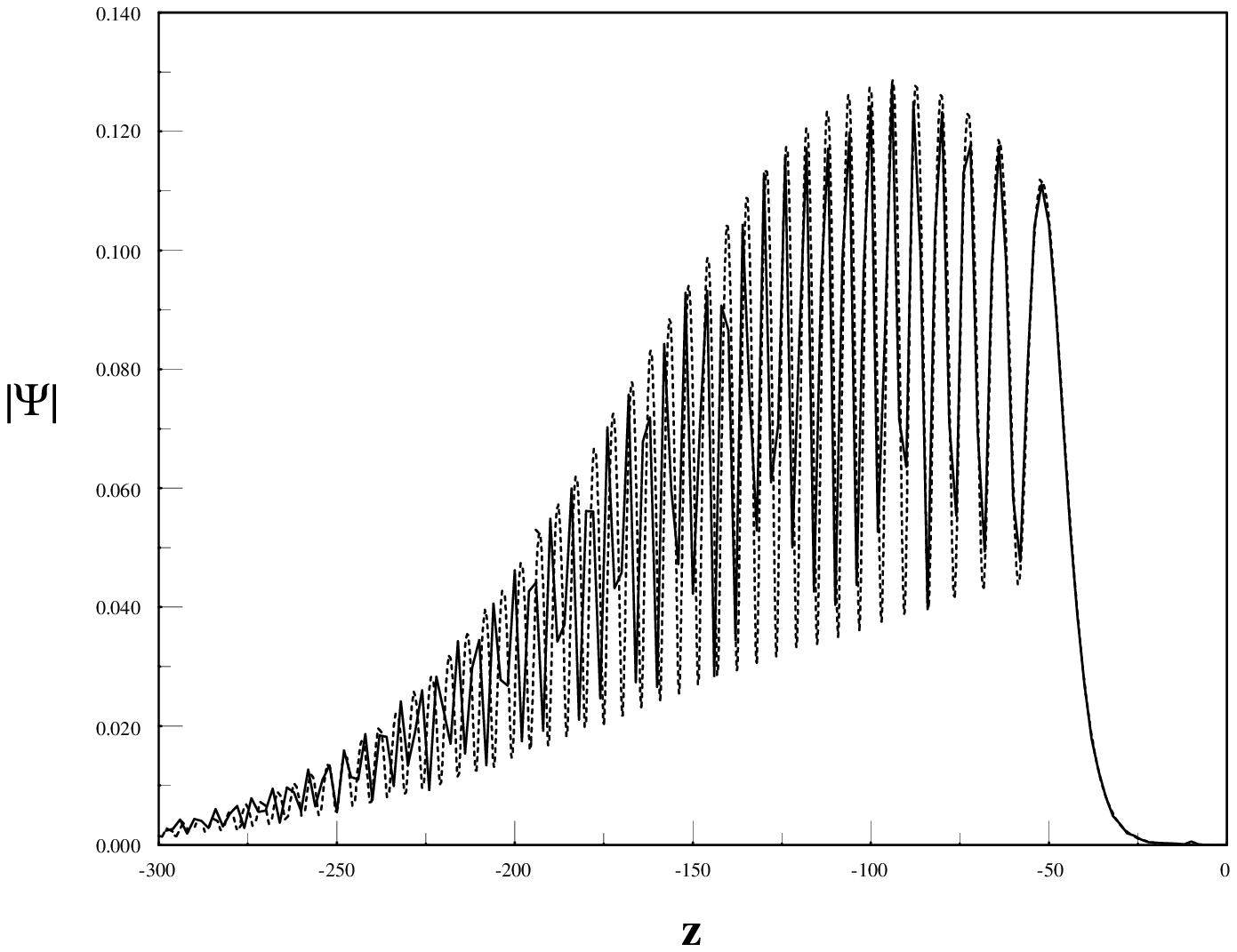}
\vsize=5 cm
\caption{\sl Numerical profile, dashed line, and,
analytical solution of eq.(\ref{expant}), solid line
 for the thin packet of figure 1}
\label{fig10}
\end{figure}

\begin{figure}
\epsffile{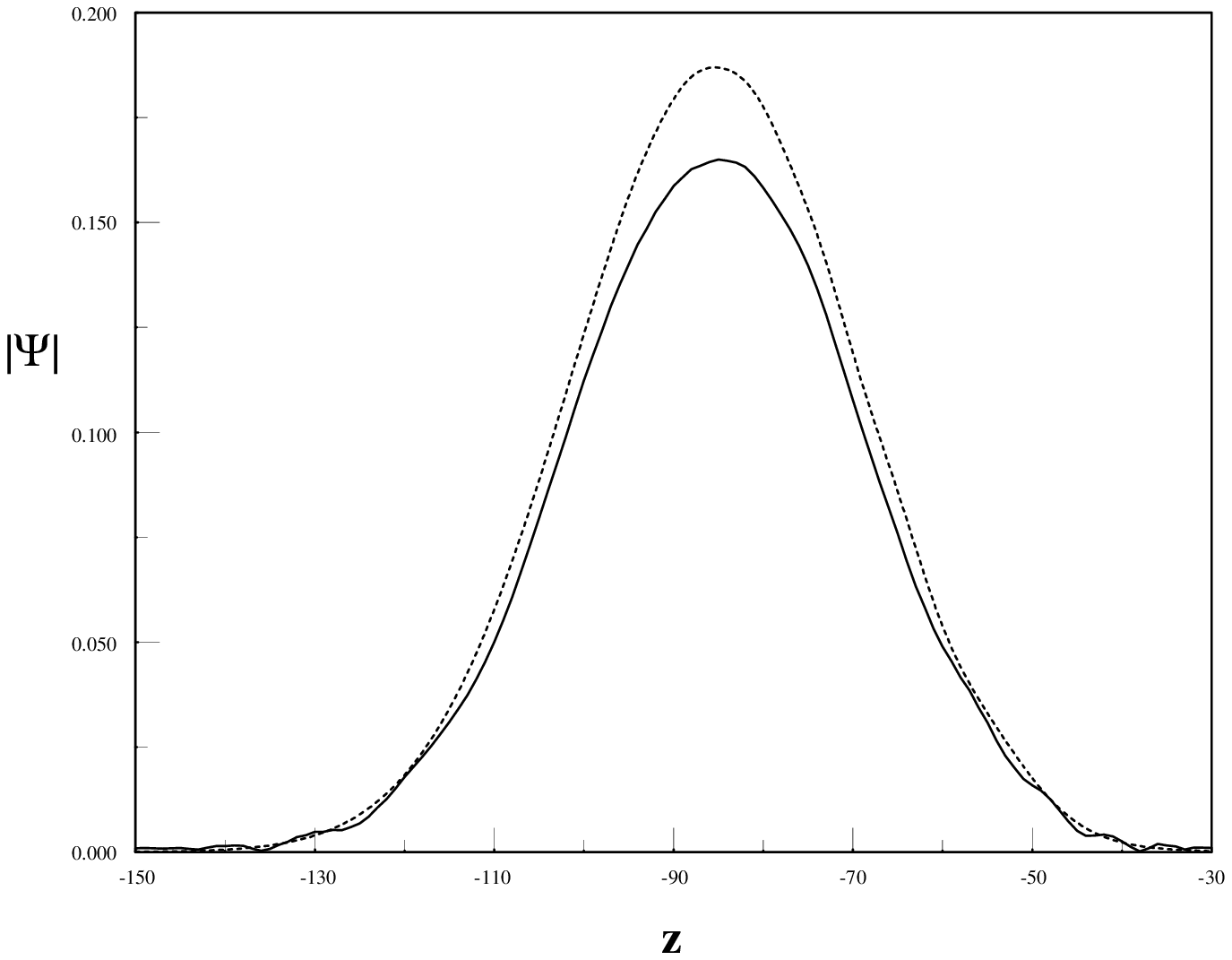}
\vsize=5 cm
\caption{\sl Numerical profile, dashed line, and,
analytical solution of eq.(\ref{expant}), solid line
 for the wider packet of figure 1}
\label{fig11}
\end{figure}
 
The long time behavior of the thin packet as compared to a wide packet remains
unchanged for as far as we could integrate the time dependent
Schr\"odinger equation numerically and still agrees
 with the analytical solution of eq.(\ref{expant}). 
We reached a time of 30 msec and
the profiles just spread, but do not change in shape.
At that time the center of the packet is around 4.5 mm below the
mirror. 

The integration of the equation becomes prohibitive beyond this
time due to the strong oscillations in the wave function that require
an increasingly smaller time step.
The results however look quite firm. The diffractive structure 
persist to infinite times, as it was found for the wavepacket
diffraction in space and time effect.\ci{k1}-\ci{k5}

In the next section we will provide some closing statements on the
relevance of the falling packet effect for Bose-Einstein condensates and
atom lasers.

 \section{\sl Summary}

We have found that a falling packet that is blocked by an obstacle
{\sl from above}, behaves differently depending on its initial spread.
A packet wider than the gravitational mass dependent length scale $\ds l_g$
falls almost as a free packet except for a lag due to an effective attraction
to the barrier or well, while a thin packet has a distinctive diffraction
pattern that propagates with it analogous to the 'Wavepacket diffraction
in space and time effect'\ci{k1}-\ci{k5}.

We consider now a possible scenario to implement the findings of this
work, and also take advantage of them for the purposes of creating
an atom laser.
Although there seems to be some controversy as to what an {\sl atom laser}
is\ci{wiseman}, at a pedestrian level it would consist in a three stage
machine: A
feeding stage that pumps in atoms in an incoherent phase; a condensation
cavity and
a continuous output coupler. The investigation of these stages is 
as of today very advanced both theoretically and experimentally.\ci{ballagh}

We would like to point out that the results of this paper suggest an 
alternative avenue for the continuous output coupling of a Bose-Einstein 
condensate.

Basically, it consists of an orifice,
a physical one or one drilled in the mesh of laser radiation that 
confines the condensate.
Through the orifice the condensate can exit in a pencil-like thin jet of atoms.
\ci{bongs}

Position now a mirror above the atoms and let them fall freely while the
feeding continues. It appears then that the outcome will be a 
coherent train of atoms having the characteristic oscillations found here,
provided the width of the pencil is smaller than $\ds \sqrt{l_g^3/z_0}$. 

In order to see if this design works and check whether it fits the
criteria of an atom laser set in ref.\ci{wiseman}, we need to perform
at least a two-dimensional calculation with a source term.
This endeavor is currently underway.

\section {\sl Acknowledgment}

I would like to thank the anonymous referees for very valuable remarks.

\end{document}